# From 2015 to 2023: How Machine Learning Aids Natural Product Analysis

Suwen Shi[1, *], Ziwei Huang[2], Xingxin Gu[3], Xu Lin[4], Chaoying Zhong[5], Junjie Hang[6], Jianli Lin[7], Claire Chenwen Zhong[8], Lin Zhang[9], Yu Li[10], Junjie Huang[8, *]

1. Department of Chemistry, Boston University, Boston, Massachusetts, 02215 United States
2. Department of Physics, Boston University, Boston, Massachusetts, 02215 United States
3. College of Professional Studies, Northeastern University, Boston, Massachusetts, 02215 United States
4. Department of Thoracic Surgery, The First Affiliated Hospital, School of Medicine, Zhejiang University, Hangzhou, Zhejiang, China
5. Department of Electrical Engineering and Automation, Guangdong Ocean University, Guangdong, China
6. Cancer Hospital & Shenzhen Hospital, Chinese Academy of Medical Sciences and Peking Union Medical College, Guangdong, China.
7. Peking-Tsinghua Center for Life Sciences, Academy for Advanced Interdisciplinary Studies, Peking University, Beijing, China.
8. The Jockey Club School of Public Health and Primary Care, Faculty of Medicine, The Chinese University of Hong Kong, Hong Kong SAR, China
9. Suzhou Industrial Park Monash Research Institute of Science and Technology, Suzhou, China
10. Department of Computer Science and Engineering, The Chinese University of Hong Kong, Hong Kong SAR, China

**Author emails:**
Suwen Shi: swshi@bu.edu, Ziwei Huang: ziwhuang@bu.edu, Xingxin Gu: gu.xingx@northeastern.edu, Xu Lin: linxu_001@zju.edu.cn, Chaoying Zhong: zhongcy3@163.com, Junjie Hang: hjj199141@alumni.sjtu.edu.cn, Jianli Lin: ljl1501@stu.pku.edu.cn, Claire Chenwen Zhong: chenwenzhong@cukh.edu.hk, Lin Zhang: tony1982110@gmail.com, Yu Li: liyu@cse.cuhk.edu.hk, Junjie Huang: junjiehuang@cuhk.edu.hk

**\*Correspondence:**
*Suwen Shi*, Department of Chemistry, Boston University; **Tel**: (857) 445 8290; **Email**: swshi@bu.edu; **Address**: 590 Commonwealth Ave # 299, Boston, Massachusetts, 02215 United States

*Junjie Huang,* The Jockey Club School of Public Health and Primary Care, Faculty of Medicine, Chinese University of Hong Kong; **Tel**: (852) 2252 8707; **Email**: junjiehuang@cuhk.edu.hk; Address: 5/F, School of Public Health, Prince of Wales Hospital, Hong Kong

**Conflict of interests:** None

**Funding:** None

**Abstract**

In recent years, conventional chemistry techniques have faced significant challenges due to their inherent limitations, struggling to cope with the increasing complexity and volume of data generated in contemporary research endeavors. Computational methodologies represent robust tools in the field of chemistry, offering the capacity to harness potent machine-learning models to yield insightful analytical outcomes. This review delves into the spectrum of computational strategies available for natural product analysis and constructs a research framework for investigating both qualitative and quantitative chemistry problems. Our objective is to present a novel perspective on the symbiosis of machine learning and chemistry, with the potential to catalyze a transformation in the field of natural product analysis.

**Introduction**

The utilization of natural products has a long history. [1] The foundational science of natural product chemistry can be tracked back to the isolation of morphine from opium in the early 19th century by Sertürner, who was the founder of alkaloid research. [2] The development of isolation and characterization of natural products accelerated in the 20th century with the advent of instrumental analysis techniques, such as nuclear magnetic resonance spectroscopy (NMR) and mass spectrometry (MS).[3,4] In the 21st century, owing to advances in organic synthesis and genomics, the impact of natural products in drug discovery, drug development, green chemistry, etc[5,6] has significantly increased. Recent research shows that from 2000 to 2014, 34 ± 9% of small molecules approved each year by the U.S. FDA(and similar organizations) can be traced to natural products or direct derivatives. [7]

Along with the development of chemical and biological techniques, the application of machine-learning methods provides more efficient and accurate innovative probabilities, including but not limited to molecule design[8], reaction prediction[9], quantum chemistry[10], material science[11], bioinformatics and genomics[12]. In natural product analysis, machine learning methods also offer significant advantages. For example, traditional spectroscopic techniques used for dereplication such as GC-MS and NMR, can be time-consuming and may have low accuracy in distinguishing similar compounds. Machine learning classification models designed for spectral analysis can expedite this process by quickly matching patterns in spectral data[13]. Furthermore, predicting metabolic pathways using traditional biochemical methods is often complex and slow. In this case, machine learning models, decision trees for example can be trained on extensive biochemical datasets to predict pathways with much higher speed and precision, circumventing the complexity and resource-intensive nature of traditional biochemical methods[14].

Sophisticated analytical instruments play a pivotal role in advancing the field of natural product analysis, falling under three overarching categories. High-Performance Liquid Chromatography (HPLC) [15], UV-Vis spectroscopy[15], and Nuclear Magnetic Resonance (NMR)[17] are employed primarily for quantitative analysis, while MS and Infrared Radiation (IR) spectroscopy[16,17] serve the purpose of quantitative evaluation. In addition to these methods, hyphenated techniques like LC-MS and GC-MS find application in semi-quantitative analysis[18]. Each of these chemical methods comes with distinct theoretical limitations. For instance, HPLC, known for its efficiency, speed, and selectivity in separating mixture components, necessitates precise solvent selection for the mobile phase and meticulous temperature control[19]. IR spectroscopy offers a wealth of information, but its information density can pose challenges in quantitative analysis and precise correlation.[20].

Despite the difficulty in breaking through the limitations of these methods from a chemical perspective, machine learning provides new possibilities. The application of machine learning (ML) techniques to chemical problems is yet to be clearly standardized. While several review articles delve into the use of ML models in this field, they often concentrate on the models' deployment rather than the overarching methodology. For instance, the review by Mullowney et al.[21] discusses the utilization of targeted ML models to discern potential drug candidates from natural products, driven by specific motivations. Nevertheless, the critical aspect of establishing a standardized protocol for ML applications in chemistry remains largely unaddressed. It is this gap that this review aims to fill, proposing a structured approach to ML integration in chemical research.

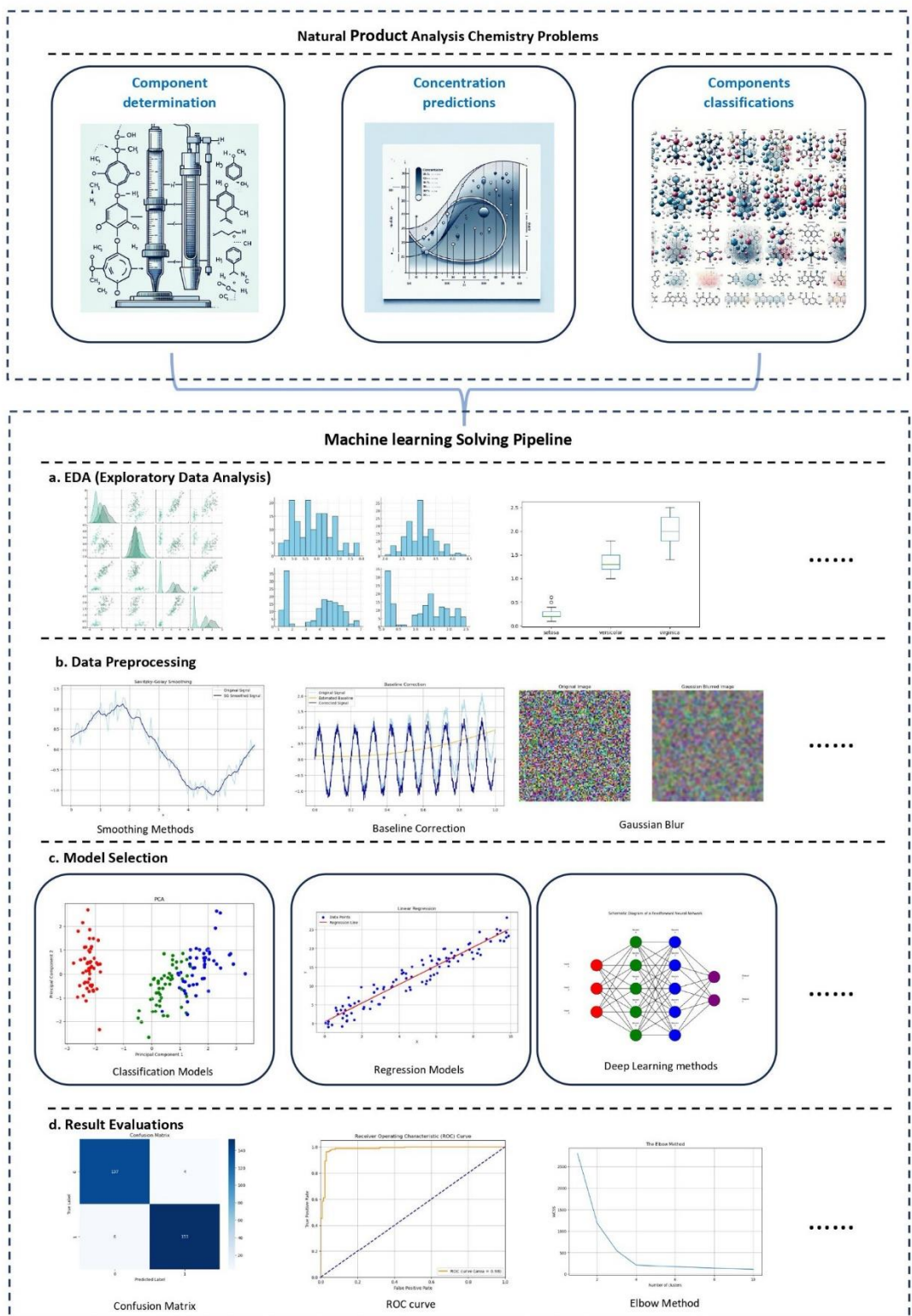

Figure 1. Machine Learning(ML) Pipeline to Solve Traditional Chemistry Problems. This figure illustrates the pivotal role of Machine Learning (ML) in enhancing the analysis of component determination, concentration prediction, and component classification in natural

product analysis. It highlights the standard ML pipeline to solve chemistry problems, which includes a) Exploratory Data Analysis (EDA), b) Data preprocessing methods (e.g., smoothing methods, baseline correction, Gaussian blur), c) Model selection (e.g., classification models, regression models, deep learning models), and d) Result evaluations (e.g., Confusion Matrix, ROC curve, Elbow Method). This integration is crucial for breaking through the limitations of traditional chemistry methods and enhancing analysis in the natural product area.

In this review, as shown in Figure 1, we provide a standard approach to match the natural product problems with machine learning methods and start with the exploratory data analysis (EDA) as the first step.
Exploratory data analysis(EDA) is fundamental for the use of machine learning techniques[22]. Some typical problems such as the need for data balance and the effects of imbalance in data sets will be explored. Progressing from the data preprocessing phase, we scrutinize three common types of data—spectroscopy data, concentration data, and image data in the natural product analysis field. Each type has unique processing steps[23], such as baseline correction, noise reduction, and Savitzky-Golay (SG) smoothing for spectroscopy data.

Next, we venture into the selection of suitable modeling techniques to tackle specific natural product problems. Model selection is a critical step and may significantly influence the outcome of the analysis. Specifically, this review matches several models with quantitative, qualitative, and semi-quantitative problems such as linear regression for the concentration prediction, logistic regression, tree-based models, partial least square (PLS), and principal component analysis (PCA) for the determination of the components, along with support vector machines (SVM) for the classification of the components.

Ultimately, the effectiveness of these models is evaluated using various result assessment methods. Our objective is to present a thorough understanding of the end-to-end process involved in natural product analysis, making it an invaluable resource for researchers and learners in the related field of chemistry.

# 1. Chemical methods for natural products analysis (problems)
## 1.1 Components determination (discover)
The analysis of natural products encompasses a transition from qualitative to quantitative evaluations. This involves initially identifying unknown components, followed by predicting their concentrations, and ultimately classifying them based on similar chemical properties or intrinsic relationships. The determination of components is the foundational step in natural product analysis, involving the identification of various individual constituents, such as botanical extracts, herbal medicines, essential oils, etc. This process includes extraction, separation, identification, and quantification. Solvent extraction, supercritical fluid extraction, and steam distillation are commonly used for extracting raw materials. Subsequently, spectroscopic techniques such as mass spectrometry (MS) and infrared (IR) spectroscopy, as well as chromatographic techniques including gas chromatography (GC), high-performance liquid chromatography (HPLC), and thin-layer chromatography (TLC), are employed for separation.
MS is often fused with other chromatographic methods (GC-MS or LC-MS) to obtain detailed molecular information[24]. For instance, Nagy et al. [25] highlighted the use of micellar electrokinetic capillary chromatography with MS in artemisinin determination. Similarly, Khajainia et al [26] showcased an efficient LC-MS/MS method for analyzing docetaxel-loaded poly nanoparticles.
However, there are challenges in interpreting MS data, especially in complex samples, and the sample preparation can be time-intensive.

IR methods, such as Fourier-transform infrared spectroscopy (FTIR), near-infrared (NIR) spectroscopy, and attenuated total reflection Fourier transform infrared (ATR-FTIR), offer insights into molecular vibrations, useful in identifying mixture components in the natural product analysis. Boukir et al.'s research [27] shows a spectroscopic study of Argan wooden artifacts which provides a strategy for an accurate estimation of the degradation level of Argan wood by evaluating the C=O bands evolution. Patle et al [28] used spectroscopy to estimate the degradation level of Argan wood, while Nugrahani et al utilized FTIR for simultaneous content determination. Although IR methods are versatile, they primarily identify functional groups without providing comprehensive structural details. Additionally, their data can sometimes be overwhelmed by background interference, which limits their quantitative analysis capabilities.

## 1.2 Concentration predictions (determination)
Predicting concentrations is crucial in natural product analysis to quantify specific compounds or groups of compounds in samples. HPLC stands out for its ability to separate, identify, and quantify mixture components. It's employed in determining ephedrine alkaloids in Ephedra products and analyzing Thiophenes in Tagetes, to name a few [29]. HPLC's high-resolution sensitivity facilitates accurate quantification of target compound. However, it requires extensive sample preparation, incurs high costs, and necessitates specialized operation.

NMR spectrometry, which offers detailed molecular structure information, utilizes parameters such as chemical shifts and peak areas for structure elucidation and component quantification[30]. Specifically, NMR aids in concentration prediction through various methods, including Quantitative 1D Proton NMR (qHNMR), standard-based quantification, quantitative 2D NMR, and metabolomic analysis [31–34]. For instance Phansalkar et al.[35] used qHNMR for total isoflavone content determination while Mansfield et al. leveraged 2D NMR for lignin study. Despite its effectiveness, NMR requires high sample concentrations and expertise in interpretation, particularly for rare products, which can be a hindrance. Advances in NMR hardware, such as low-temperature probes, are addressing some of these limitations.

UV-Vis spectroscopy aiding in quantifying light-absorbing molecules.  The Beer-Lambert Law allows for the direct quantification of concentration based on a solution's absorbance [36]. Examples include Zhu et al.'s determination of flavonoids from P. oleracea L and Martelo-Vidal et al.'s rapid quantification of red wine polyphenols. UV-Vis's utility is mostly for colored compounds, and its precision can suffer from interference in complex mixtures.

**1.3 Components Classification**
Components classification, usually regarded as qualitative or semi-quantitative problems usually refers to classifying different groups of components in a mixture system. Compared with the components determination problems that focus on one component, components classification involves sorting components based on their chemical structure (e.g., terpenes, alkaloids, phenolics), their biological activity, or other relevant properties. Some hyphenated techniques, GC-MS, LC-MS for example, FTIR, and experimental chemistry methods can be used.

MS-based techniques, such as CG-MS and LC-MS, are often employed for this task. GC-MS is often likened to an "electronic nose" due to its ability to provide a spectral fingerprint of compounds. LC-MS, on the other hand, leverages the mass-to-charge ratio (m/z) and fragmentation patterns to identify compounds. In the context of qualitative analysis, these techniques are indispensable for characterizing components in a mixture. Subsequently, the obtained data can be compared against a mass spectral library to elucidate the specific components present.

FTIR is another option. Using FTIR methods, many characteristic peaks can be identified. The qualitative analysis can be achieved by comparing it with the spectroscopy library and the quantitative analysis can be achieved by using Beer-Lambert Law. However, the result of FTIR methods usually cannot be persuasive because of the noise and the background influences. To break through this principle limitation, there are several solutions. Melucci et al.[37] introduce a non-destructive ATR-FTIR method to analyze the biogenic silica in marine sediments which is a new quantification approach. Neves et al.'s study[38] shows an ATR-FTIR method to detect adulteration with all test oils in the concentration of 10-40%.

## 2. Machine learning strategy for the natural product analysis problems
At the beginning of using machine learning strategies for natural product analysis problems, it's worth clarifying that we need to follow the machine learning pipeline. To be specific, the following steps should be followed: 1. Data collection 2. EDA 3. Data preprocessing 4. The model employment 5. Model Evaluation. For each step, specific techniques can be employed for the target.

**2.1 EDA(Exploratory data analysis)**
Exploratory data analysis (EDA) in machine learning is an approach that uses various techniques, such as visual methods, to maximize the insights gained from data[39–43]. EDA is the most fundamental step, usually being the first in machine learning analysis, responsible for data structure clarification, model selection, potential errors, outlier identification, and more. It provides a great opportunity for natural product researchers to develop hypotheses and topics for further analysis.
EDA usually involves the following activities:
1) Summary statistics[44]
   Summary statistics involve calculating central tendency (mean, median, mode), measures of dispersion (range, interquartile range, variance, standard deviation), and measures of data shape (skewness, kurtosis). One typical problem is the data balance problem. Especially in natural product analysis, the predictive model performance may be biased toward the class with a large number of samples.
2) Data cleaning
   The data cleaning step varies based on different types of natural product data. Typical data cleaning includes imputation (replacing missing values with mean, median, or mode), missing value deletion, and missing value prediction.
3) Visualization
   Visualization techniques are an intuitive way to express data distribution, trends, and relationships. Visualization techniques include histograms, box plots, scatter plots, heatmaps, bar charts, and more. For example, in natural product analysis, a heatmap[45] is a good method to show the basic relationship between the selected variables.
4) Testing assumptions
   This step of the EDA is based on the model chosen. For instance, using statistical tests like the chi-square test can help confirm independence in the linear regression model.
5) Feature binning, encoding, and dimension reduction[46].
   Feature engineering is a significant topic that includes encoding, standardization, and interaction features. The primary purpose of feature selection is not to maximize the accuracy of the models but to improve their efficiency. The feature engineering process is essential throughout the entire data analysis and cannot be easily categorized into EDA or Data preprocessing parts. In this review, we roughly place feature engineering into EDA and summarize common techniques used in feature engineering:
   A. Binning[47]: In spectroscopic data, a simple approach is to regard each wavenumber as one feature and convert continuous numerical features into categorical counterparts like "Peak 1," "Peak 2," etc. Common practice includes setting a bin value as the standard. Other techniques, such as wavelet transform or auto-binning, have also developed as alternatives. In Sankaran et al.'s research[48] on huanglongbing detection based on visible-near infrared spectroscopic data, they performed data binning by averaging normalized spectra for each 25 nm wavelength, reducing the features from 989 to 86.
   B. Category encoding: Category encoding converts categorical data into numerical data. For example, one-hot encoding is widely used in machine learning to convert each categorical value into a new categorical column, assigning a binary value of 1 or 0. In Lv et al.'s work[49], they use a dinucleotide one-hot encoder to transfer DNA into tensors and achieve a test accuracy of 96.19%.
   C. Dimension reduction[50–52]: In natural product analysis, dimensionality reduction simplifies large and complex datasets, such as mass spectrometry, NMR, and various chromatographic techniques. Principal Component Analysis (PCA) and Partial Least Squares Discriminant Analysis (PLS-DA) are popular techniques. PCA is an unsupervised learning method used to identify the most significant spectra or frequency bands contributing to the dataset's variance. PLS-DA is a supervised learning method, a variant of PCA, used to separate components into categories. Mohamad Asri et al.[53] introduced a novel approach for dimension reduction by using PCA, and the PLS-DA model was also trained to reduce the number of classes, achieving a 91% classification rate.

## 2.2 Data preprocessing
### 2.2.1 Spectroscopic data

Spectroscopic data is the result of employing diverse spectroscopic techniques, typically yielding spectra comprising attributes like wavenumbers, absorptions, and intensities. Such data is inherently characterized by high dimensionality, organized in a matrix structure where individual rows pertain to distinct samples, while each column, designated as a feature or variable, signifies a spectral wavelength or frequency. Within each cell of this matrix, the value represents the intensity or absorbance of light associated with the corresponding wavelength.

The fidelity of a spectrum's baseline is frequently compromised by various factors, including instrumental noise, scattering effects, and sample heterogeneity. Consequently, it becomes imperative to rectify the baseline before embarking on further analytical procedures. This correction can be achieved through a range of methods, such as Asymmetric Least Squares (ALS), Polynomial Fitting, and Iterative Techniques. As shown in Table 1, we designate the "Asymmetric Least Squares (ALS)" method as the established standard and compare the other methods with it. ALS is widely acknowledged for its versatility and robustness in addressing intricate baselines encountered in analytical chemistry, owing to its proficient management of asymmetrical deviations and noise.

Table 1A. Approach and Complexity

| Name of the Method | Basic Approach | Complexity |
|---|---|---|
| ALS (Standard) | Uses weighted penalties for a smooth baseline curve | High - Handles complex baselines |
| Polynomial Fitting | Fits a specified order polynomial to the baseline | Lower - Effective for simpler baselines |
| Iterative Methods | Subtracts baseline iteratively using a smoothing method | Variable - Depends on specific method |

Table 1B. Performance and Implementation

| Name of the method | Robustness to Noise | Computation Time | Assumptions | Implementation Difficulty |
|---|---|---|---|---|
| ALS (Standard) | Robust with appropriate penalty weights | Higher due to optimization | Assumes positive deviations are more probable | Medium - Requires tuning |
| Polynomial Fitting | Less robust, prone to overfitting/underfitting | Lower for low-order polynomials | Assumes baseline is a polynomial | Easier - Common statistical technique |
| Iterative Methods | Variable depending on specific method | Variable | Variable depending on method | Variable depending on method |

Table 1C. Suitability and Limitations

| Name of the method | Suitable Use Cases | Drawbacks |
|---|---|---|
| ALS (Standard) | Complex baselines, asymmetric deviations expected | Requires selecting appropriate penalty weights |
| Polynomial Fitting | Simple baselines that can be approximated by a polynomial | Prone to overfitting or underfitting with incorrect order |
| Iterative Methods | Complex baselines, iterative refinement enhances correction | Can be slow and computationally intensive |

Normalization[54]: Normalization is a commonly used method to adjust values to a common scale. Normalization usually involves several methods, as shown below:
1) Scaling: Scaling is the simplest normalization method. The function is (spectrum - min(spectrum)) / (max(spectrum) - min(spectrum)), which scales the values between 0 and 1.
2) Standard Normal Variate (SNV): This method is usually employed to normalize the whole spectrum to the intensity of a specific constant peak in several samples. The SNV method has been widely used to normalize multiple spectra measured from one set of data samples. Bi et al.[55] introduce that SNV transformation reduces the multiplicative effects of scattering by scaling and centering the spectrum with standard deviation, improving the linearity of PC1 (the first component of PCA) from $r = 0.55$ to $r = 0.77$.
3) Z-score Normalization: -score can be expressed as: $Z = (X_i – X_{avg})/SD$. It is usually used to standardize the range of the independent variables or features of data. Zeng et al.[56] show that the accuracy of the model trained with z-score normalization is higher than 96% in LSTM, BILSTM, and other models in a neural network.

Noise Reduction: Noise reduction is one of the most important parts of spectroscopy data preprocessing. For example, IR spectroscopy includes a lot of information, which often results in significant background noise. Data preprocessing methods help to reduce noise from the instrument, environment, random errors, etc. Here are some methods for spectroscopy data in natural product analysis:

1) Savitzky-Golay smoothing[57]: This method applies a polynomial filter to the data to smooth and reduce noise. The performance is shown in Figure 1. Under this process, the peaks of the spectra are preserved to some extent. SG-smoothing has been employed in Ning et al.'s research to improve smoothness and reduce noise (Figure 2). They achieved a root mean square error of prediction of $2.1 \mu g \cdot kg^{-1}$ with the LASSO-SVM model.

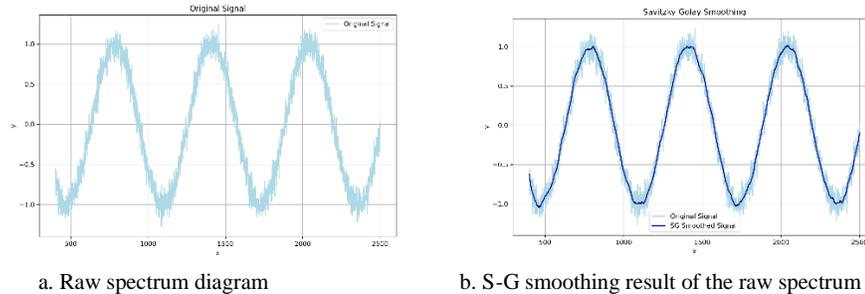

a. Raw spectrum diagram          b. S-G smoothing result of the raw spectrum

Figure 2: Comparison of the raw spectrum and the S-G smoothing spectrum. Figure 2a. presents the raw spectroscopic data, characterized by noise. Figure 2b. illustrates the raw spectra subjected to the Savitzky-Golay (SG) smoothing methods, The SG smoothing technique enhances the spectral smoothness and mitigates noise.

2) Wavelet Transform[46]: Wavelets can decompose a signal into different frequency components. Wavelet transform can be used to decompose and reconstruct the signal which is very useful to denoise and reduce the feature. Li et al.[59] apply lifting wavelet transform (LWT) to improve the $R^2$ of the calibration set from 0.749 to 0.809. Agustika et al.[60] apply level five discrete wavelet transform (DWT) to the reduce raw spectrum from 3734 data to 117.

3) Multiplicative Scatter Correction (MSC): MSC and related methods strongly help to correct scatter effects and baseline shifts in spectroscopy data. Mishra et al.[61] improved prediction results by using scatter correction techniques. Compared with SNV-corrected data in their research, the prediction error was reduced by up to 15% and 57%.

4) Principal Component Analysis (PCA): PCA is an unsupervised learning algorithm usually used for dimension reduction. In the case of noise reduction, PCA detects the most significant components with the largest variance. Kusaka et al. employed PCA to reduce the noise in solid-state NMR spectra and discarded 95 noise factors in one of the experiments, which is significant.

Here, we designate the S-G smoothing technique as the reference standard for noise reduction owing to its straightforward implementation, minimal complexity, and efficacy in smoothing and noise reduction within regularly spaced data sets.

Table 2. Comparison of common noise reduction models.

Table 2A. Approach and Complexity Comparison

| Name of the Method | Basic Approach | Complexity |
|---|---|---|
| S-G Smoothing (Standard) | Uses a moving window and polynomial fitting to smooth the data | Lower - Easy to implement |
| Wavelet Transform | Transforms the data into a different domain where noise can be separated from signal | Higher - Requires a good understanding of wavelets |
| MSC | Corrects scatter effects to standardize samples | Medium - Involves calculation of correction coefficients |
| PCA | Reduces the dimensionality of the data, capturing most of the variance in fewer dimensions | Higher - Involves eigenvalues and eigenvectors |

Table 2B. Performance and Implementation Comparison

| Name of the Method | Robustness to Noise | Computation Time | Assumptions | Implementation Difficulty |
|---|---|---|---|---|
| S-G Smoothing (Standard) | Robust for certain types of noise, depending on window size and polynomial order | Lower - Depends on window size and polynomial order | Assumes noise is high frequency compared to signal | Easier - Many libraries available |
| Wavelet Transform | Highly robust to noise, particularly non-stationary noise | Higher - Depends on wavelet and level of decomposition | Assumes noise can be separated from signal in wavelet domain | Medium - Requires understanding of wavelets |
| MSC | Less robust to noise, primarily intended to correct scatter | Medium - Involves a simple calculation | Assumes scatter is multiplicative and is the main source of variation | Easier - Many libraries available |
| PCA | Highly robust, can separate noise from signal | Higher - Involves matrix operations | Assumes noise can be separated from signal in PCA space | Medium - Requires understanding of PCA |

Table 2C. Suitability and Limitations Comparison

| Name of the Method | Suitable Use Cases | Drawbacks |
|---|---|---|
| S-G Smoothing (Standard) | Works well for smoothing and noise reduction in evenly spaced data | Can distort signal if window size or polynomial order is not chosen correctly |
| Wavelet Transform | Works well for non-stationary signals and complex noise | Can be complex and require fine-tuning |
| MSC | Works well for scatter correction in spectral data | Only corrects scatter, not other types of noise |
| PCA | Works well for high-dimensional data where noise and signal can be separated | Requires sufficient data for PCA, can be complex to understand |

Alignment methods are groups of methods to deal with multiple sample data. For example, COW is a method of stretching and compressing the spectral axis to achieve the best correlation. Kucha et al.[62] indicate that COW is effective in near-infrared hyperspectral imaging which allows the high accuracy($R^2 = 0.89$) of support vector machine regression. Some other methods, Peak Alignment by Fast Fourier Transform (PAFFT) or iterative closest point can also be used. Sun et al.[63] illustrate PAFFT to deal with the peak shift problem in NMR data, which consumed less than 30 seconds with high efficiency.

#### 2.2.2 Concentration data and image data preprocessing

In addition to spectroscopic data, concentration data and image data should also be considered. Concentration data, derived from tools such as HPLC and LC-MS, provides insights into the components of samples in natural product analysis. For optimal machine learning training, normalization techniques, such as Z-score normalization, and transformation methods, such as log transformation, are essential. Concentration data often synergizes with spectroscopic data, as seen in Chen et al.'s work integrating HPLC and infrared data. On the other hand, image data, interpreted as pixel data, originates from sources like microscopy and IR imaging. Key preprocessing techniques include image normalization, ensuring pixel values between 0 and 1, as illustrated by Lu et al.'s hyperspectral imaging study. Additionally, data augmentation expands datasets by introducing variations like flips and rotations, enhancing model performance. Ferentinos et al.'s resizing approach is a prime example, improving efficiency to 90.79%. Noise reduction, encompassing methods from Gaussian blur to advanced strategies like denoising convolutional neural networks (DnCNNs), is crucial to refine image data quality.

## 2.3 Model selection for natural products problems

Upon completing the EDA and data preprocessing steps, finding the most suitable machine learning models will directly influence the results. Basically, once the model has been selected, all of them deserve a hyperparameter tuning part at the beginning which will be very helpful to improve the performance of the model. Tuning includes pre-tuning and fine-tuning. Pre-tuning involves setting a wide range for the models to ensure the selected model fits the data with no obvious errors. Then, data needs to be split into the train set and the test set. Some feature selection models can be used on the train set to reduce features for the fine-tuning process. Once the best parameters have been determined, the test set can be used to verify the performance of the model and ensure there is no overfitting or underfitting problem. The prediction process will run after the model has been established.

For the model selection step, it's necessary to categorize the data into numerical and categorical at the beginning. Based on different data types, different models can be chosen. Supervised learning models are trained using labeled data and aim to map the input data to the known output. Typical supervised models involve regression models and classification models. Unsupervised learning models, compared to supervised learning models, are trained using unlabeled data and aim to generate unknown outputs. Unsupervised learning models can be divided into clustering models and dimensionality reduction models.

It's worth pointing out that in this review, in natural product analysis, we group the classification models and clustering models as "classification models." Moreover, we will not include dimensionality reduction models as a separate group because these models do not appear as a complete analysis process in the analysis of natural product data but as a step.

#### 2.3.1 Models for concentration prediction

Concentration prediction problems, serving as quantitative analysis problems, match well with regression models in machine learning. Regression models estimate the relationship between a dependent variable (concentrations, for example) and one or more independent variables. Commonly used regression models for concentration prediction include:

1) Linear Regression

    Linear regression is a simple machine learning model that predicts concentration based on the linear relationship among the input features. In natural product analysis, multiple features commonly exist. Multiple linear regression, based on multiple features, is widely used. Meina et al[64] introduced a multiple linear regression approach to predict the correlation between how chemistry variables influence toxicity to organisms, such as pH changes.

2) Ridge Regression

    Compared to simple regression models, Ridge Regression adds a penalty term to the loss function to avoid the overfitting problem caused by high correlation among features. Singhal et al.[65] estimated leaf chlorophyll concentration using Ridge

regression and achieved a Root Mean Squared Error (RMSE) of 0.10 mg/g and R² = 0.7452, which is a good attempt but not good enough.

3) Lasso Regression

Similar to Ridge regression, Lasso regression can be employed for feature selection. Generally, Lasso regression adds L1 regularization to the loss function of the least squares method to shrink the coefficients of less important features close to zero. Xu et al.[66] employed and compared LASSO and RT methods in generic E.Coli concentration prediction and achieved an RMSE of 0.974 with LASSO, providing a practical tool to predict E.coli population.

4) Elastic net

Elastic net is a powerful linear regression that essentially combines Ridge Regression and Lasso Regression. Elastic net can handle both multicollinearity and feature selection problems. Zeng et al.[67] employed the non-negative LASSO (NN-EN) algorithm to identify all the components in the mixture and achieved high-accuracy components identification and low-error concentration prediction.

Here, Multiple Linear Regression has been designated as the standard methodology for the analysis of natural product concentration prediction due to its fundamental approach and inherent simplicity, making it a fundamental form of regression analysis.

Table 3. Comparison of Multiple Linear regression, Ridge Regression, Lasso Regression, and Elastic Net

Table 3A. Approach and Complexity Comparison

| Name of the Method | Basic Approach | Complexity |
|---|---|---|
| Multiple Linear Regression (Standard) | Minimizes the sum of the squared differences between the observed and predicted values | Lower - Basic form of regression analysis |
| Ridge Regression | Adds a penalty equivalent to the square of the magnitude of coefficients | Medium - Due to the introduction of regularization |
| Lasso Regression | Adds an absolute value of magnitude of coefficient as penalty term | Medium - Due to the introduction of regularization |
| Elastic Net | A combination of Ridge and Lasso, uses both L1 and L2 as penalty term | Higher - Combines aspects of both Ridge and Lasso |

Table 3B. Performance and Implementation Comparison

| Name of the Method | Flexibility | Robustness to Overfitting | Computation Time | Assumptions | Implementation Difficulty |
|---|---|---|---|---|---|
| Multiple Linear Regression (Standard) | Lower - Does not include any regularization | Lower - Does not include any penalty to prevent overfitting | Lower - Does not include regularization | Assumes linear relationship, no multicollinearity, homoscedasticity, independence of errors | Easier - Straightforward with many available tools |
| Ridge Regression | Higher - Regularization can be controlled with lambda | Higher - L2 penalty reduces overfitting | Higher - Includes regularization | Same as multiple linear regression, but more robust to multicollinearity | Medium - Requires selection of lambda |
| Lasso Regression | Higher - Regularization can be controlled with lambda | Higher - L1 penalty reduces overfitting and performs feature selection | Higher - Includes regularization | Same as multiple linear regression, but more robust to multicollinearity | Medium - Requires selection of lambda |
| Elastic Net | Highest - Controls the mix of L1 and L2 regularization with alpha | Highest - Combines penalties of both Ridge and Lasso | Highest - Includes both L1 and L2 regularization | Same as multiple linear regression, but more robust to multicollinearity | Harder - Requires selection of lambda and alpha |

Table 3C. Suitability and Limitations Comparison

| Name of the Method | Suitable Use Cases | Drawbacks |
|---|---|---|
| Multiple Linear Regression (Standard) | Works well when data is linearly related and there's no multicollinearity | Due to the inherent theory, the performance is limited |
| Ridge Regression | Works well when there's multicollinearity and you want to prevent overfitting | May not perform well if multicollinearity is not an issue |
| Lasso Regression | Works well when you want to do feature selection while preventing overfitting | May pick only one variable from a group of correlated variables |

Various models are compared in Table 3. In brief, the utilization of regression methodologies—namely, Multiple Linear Regression, Ridge Regression, Lasso Regression, and Elastic Net—presents inherent limitations on concentration prediction. Multiple Linear Regression oversimplifies intricate relationships by presuming strict linearity. Ridge and Lasso Regression, despite their effectiveness in addressing multicollinearity and facilitating feature selection, necessitate meticulous parameter tuning and can encounter challenges with correlated or noisy data. Introducing complexity in parameter selection, Elastic Net amalgamates the merits of Ridge and Lasso. Furthermore, all these approaches operate under the assumption of a certain level of linearity, potentially constraining their capacity to apprehend intricate nonlinear associations within the data. Hence, practitioners must remain mindful of these constraints and judiciously select the most suitable methodology aligning with their specific dataset and analytical imperatives.

Despite the drawbacks, they all facilitate breakthroughs by enhancing accuracy, speed, and efficiency in several ways:
1) Reduce the sample size[68–70]: Regression models can achieve concentration prediction using limited physical samples, which can greatly assist in time-consuming chemistry methods such as HPLC and NMR.
2) Improve accuracy[70–72]: Traditional concentration prediction pipelines, for example, use a few data points to fit a linear line, which has great limitations in accuracy and can be difficult to deal with complex concentration data. Regression models are capable of handling high-dimensional and non-linear data.
3) Feature selection[73,74]: Regression models like Lasso and Ridge are capable of feature selection. Using these models, researchers can find the most important features strongly related to concentration prediction.

**2.3.2    Models for components determination (discover)**
Components determination problems, serving as qualitative analysis problems, can be roughly matched with classification models in machine learning. Based on different types of data, labeled or not, related machine learning models can be divided into supervised learning models and unsupervised learning models. The main topic of components determination problems is to match the unknown component to a known database. By evaluating the accuracy of the prediction result, we can answer "Which component is this?" or "How similar is this component to the known one?"
Unsupervised learning methods:
1) Principal Component Analysis (PCA): PCA is a powerful dimension reduction method that helps us to identify the key variables in a multidimensional dataset. Golimowski et al.[75] employed PCA to verify that there's no influence of bleaching earth on bleached oils, with the first principal factors accounting for 76.2% of the variation.
2) K-means K-means is a typical unsupervised learning method used to partition data into clusters. Mohanty et al.[76] used K-means clustering to understand the similarity and relationship among Hedychium species, successfully classifying samples into four major clusters.

Supervised learning methods:
1) Support Vector Machine (SVM): SVM can be used for both classification and regression tasks, by changing the kernels[77–79]. SVM models can achieve non-linear classification by finding an optimal hyperplane to separate data of different classes. This helps perform well for high-dimensional data and reduces overfitting. Sampaio et al.[80] identified rice flour types using SVM models, achieving 91% prediction accuracy after scatter correction with MSC, verifying the model's suitability for rice type classification.
2) Ensemble models: Ensemble models are machine learning models which combine multiple individual models. Those models usually provide high accuracy and robustness[81–84]. Examples include Decision Tree, Random Forest, and XGBoost, capable of complex data analysis. Fuente et al.[85] analyzed 8000 industry hemp samples using random forest classifiers and successfully explored the relationship between terpenoid content, cannabinoid composition, and subjective reports.
3) Partial Least Squares and derivative method: PLS is a statistical technique popular in natural product analysis. Some infrared instruments have this model built in to help users get a rough glance of the components. PLS and derivative methods, such as PLS-DA, are widely used in natural product analysis to analyze spectroscopy and concentration data. Walkowiak et al.[86] generated ATR-FTIR and PLS-DA methods to detect potential adulterants like kaempferol, quercetin, and rutin in their samples, achieving accuracy rates of 87.5%, 93.7%, and 87.5%, respectively.

Principal Component Analysis (PCA) has been designated as the standard technique due to its fundamental principle of dimension reduction, rendering it highly suitable for managing high-dimensional data and widely used to extract pertinent insights from intricate datasets.

We can briefly compare these models in Table 4.

Table 4. Comparison of PCA, K-means, SVM, Ensemble models and PLS
Table 4A. Approach and Complexity Comparison

| Method | Basic Approach | Complexity |
| --- | --- | --- |
| PCA (Standard) | Dimension reduction using principal components | Involves eigenvalue or singular value decomposition |
| K-means | Cluster minimization of within-cluster variance | Simple algorithm with iterative updates of cluster centers |
| SVM | Hyperplane separation of classes in feature space | Complex, involves quadratic programming and kernel selection |
| Ensemble Models | Combination of models to improve prediction | Depends on combination of models, some need extra tuning |
| PLS | Regression using latent variables combined from features | Handles multicollinearity by combining PCA and multiple regression |

Table 4B. Performance and Implementation Comparison

| Method | Flexibility | Robustness to Overfitting | Assumptions | Suitable Use Cases | Drawbacks |
|---|---|---|---|---|---|
| PCA (Standard) | Lower, determined by explained variance | Unsupervised, doesn't overfit | Linear relationships among variables | High-dimensional data, dimension reduction | Captures only linear relationships |
| K-means | Requires known number of clusters | Controlled by suitable cluster numbers | Spherical, balanced clusters | Clustering unlabeled data | Sensitive to initial values, outliers, cluster numbers |
| SVM | Decision boundaries determined by kernel selection | Controlled by penalty parameter and kernel selection | Separability of classes | Binary and multiclass classification | Kernel and parameter sensitivity |
| Ensemble Models | Applicable to various regression and classification problems | Generally robust | Depends on base models | Poorly performing or overfitting single models | Computationally expensive, hard to interpret |
| PLS | Applicable to various regression and classification problems | Controlled by selecting right component numbers | Linear relationships among variables | Highly collinear independent variables | Poor performance for nonlinear relationships, Linear relationships |

Table 4C. Suitability and Limitations Comparison

| Method | Suitable Use Cases | Drawbacks |
|---|---|---|
| PCA (Standard) | High-dimensional data, dimension reduction | Captures only linear relationships |
| K-means | Clustering unlabeled data | Sensitive to initial values, outliers, cluster numbers |
| SVM | Binary and multiclass classification | Kernel and parameter sensitivity |
| Ensemble Models | Poorly performing or overfitting single models | Computationally expensive, hard to interpret |
| PLS | Highly collinear independent variables | Poor performance for nonlinear relationships, Linear relationships |

### 2.3.3 Models for classification

Component classification encompasses qualitative and semi-quantitative analysis, which can be addressed using classification models and clustering methods. Compared with component determination problems, classification problems focus more on grouping components in a mixture system, for example, ground cannabis samples. The classification results obtained by machine learning are likely to differ from traditional chemical, botanical, or biological classifications.

Those methods bring a special angle for the researchers to break through the limitation of traditional chemistry methods not only in the principal aspect but also in the experimental aspect.

For example, feature selection and regression models make the quantitative analysis of infrared spectroscopy possible. Dimension reduction methods, PCA for example are very capable of handling data with many features. To be specific, Gambardella et al.[87] employed PCA to reduce the sample size and reinterpret the observations. By using this method, image classification has been achieved to monitor the cultivation of cannabis plantations.

Using machine learning models to classify and determine components in a mixture overcomes the theoretical limitations of chemical techniques in the following ways:
  (1) Classification limitation
      Traditional chemical classific=-ation and determination techniques are most commonly based on chromatography techniques, such as HPLC. By using PCA, PLS, and other classification models, classification problems can be easily solved through spectroscopic data with high accuracy.
  (2) Quantification limitation
      Some spectral techniques, like infrared spectroscopy, have significant limitations in quantification analysis due to high background noise. Machine learning models provide a novel approach to denoise spectra, filter features, and correct the baseline. These processes significantly aid in achieving quantification analysis, potentially breaking through the limitations of infrared spectroscopy and related techniques.

## 2.4 Result evaluation methods for problems above
### 2.4.1 Evaluation Methods for Concentration Prediction

The task of predicting concentration in natural product analysis is essentially a regression problem. In other words, our goal is to find the optimized function where the dependent variable is concentration. The commonly used metrics for evaluation include:

Mean Absolute Error (MAE): MAE is widely used for evaluating regression models[88–92] which formula is:
$$MAE = \left(\frac{1}{n}\right) * \Sigma|y - \hat{y}|$$
Where:
n is the number of samples or observations.
y is the actual value of the target variable.
ŷ is the predicted value of the target variable.
MAE provides the average magnitude of the errors made by the model. However, MAE is not as sensitive to outliers compared to MSE, and it treats overestimations and underestimations equally without providing the distributions of the errors. For instance, Nuapia et al.[93] employed multilinear regression to achieve cannabinoid concentration prediction and evaluated the model using the MAE metric with values of 0.10825 for THC, 0.005083 for CBN, and 0.029333 for CBD.

Mean Squared Error (MSE) or Root Mean Squared Error (RMSE):
Similar to MAE, MSE and RMSE are commonly used for evaluating regression models[94–97]. The formula for MSE and RMSE are:
$$MSE = \left(\frac{1}{n}\right) * \Sigma(y - \hat{y})^2$$
$$RMSE = \sqrt{MSE}$$
Where:
n is the number of samples or observations.
y is the actual value of the target variable.
ŷ is the predicted value of the target variable.
These metrics emphasize large errors more than MAE and are sensitive to outliers. For instance, Zhu et al.[98] introduced a rapid method for fatty acid composition determination, achieving an RMSE for the prediction lower than 0.947, showing great performance.
R-Squared (Coefficient of Determination):
R-squared is a value between 0 and 1 used to assess the goodness of fit for a regression model[99–102]. The R-squared value directly shows the fit result though it cannot precisely indicate the prediction accuracy. This means that we cannot ignore the overfitting problem even if the model shows good fitting performance. Turhan et al.[103] employed linear regression models to determine the relationship between 3 terpene classes and 11 cannabinoid components, achieving an R-squared value of 99.5%, showing great performance.

### 2.4.2 Evaluation methods for Components determination
Components determination involves several commonly used evaluation methods:

Silhouette Coefficient: This metric evaluates the similarity of an object to its own cluster compared to other clusters in unsupervised machine learning tasks[104–106]. The silhouette ranges from −1 to +1, where a coefficient close to +1 indicates the sample is well clustered, close to 0 indicates the data is close to the boundary, and close to -1 indicates the sample is wrongly clustered. However, the Silhouette Coefficient is not applicable to data that cannot be naturally clustered. When optimizing this value, it's important to weigh intra-cluster similarity and inter-cluster dissimilarity to improve the Silhouette coefficient. Chambers et al.[107] applied PCA to distinguish hemp and marijuana, determining that the optimal number of clusters should be two based on the silhouette method.

The Elbow method is a graphical technique to evaluate the optimal number of clusters in unsupervised learning[108–110] For example, for K-means clustering, the Elbow method plots the explained variation as a function of the number of clusters and picks the elbow of the curve as the number of clusters to use. However, the Elbow method can sometimes be challenging to use for determining the optimal number of clusters because the elbow point may not be clearly defined. Stack et al.'s research[147], the number of clusters (n=4) was determined by the elbow method for the k-means clustering.

### 2.4.3 Evaluation methods for Components classification
The commonly used evaluation methods for classification models include:
A confusion matrix provides a comprehensive view of the performance of a classification model[111–113]. It includes the following definitions:
True positives (TP): Predicted positive and are actually positive.
False positives (FP): Predicted positive and are actually negative.
True negatives (TN): Predicted negative and are actually negative.
False negatives (FN): Predicted negative and are actually positive.

Based on the confusion matrix, several other metrics can be derived:
Accuracy: The equation is: $\frac{TP+TN}{TP+FP+TN+FN}$
Precision: The equation is: $\frac{TP}{TP+FP}$
Recall/sensitivity/True-positive Rate: The equation is $\frac{TP}{TP+FN}$
Specificity: The equation is $\frac{TN}{FP+TN}$
F1 score: the equation is $\frac{2*Precision*Recall}{Precision+Recall}$

In Lu et al.'s study[114], a confusion matrix was employed to evaluate the discrimination of floral samples into rich or poor in CBD and THC.

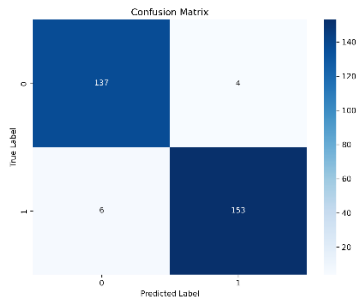

Figure 3. Example of the Confusion Matrix: This confusion matrix visualizes the performance of a binary classifier. The x-axis represents the predicted class labels, and the y-axis represents the true class labels, with '0' denoting the negative class and '1' the positive class. The matrix is divided into four quadrants: true negatives (TN) in the bottom-left, false negatives (FN) in the top-left, true positives (TP) in the top-right, and false positives (FP) in the bottom-right.

Receiver Operating Characteristic Curve (ROC Curve): The ROC curve shows the diagnostic ability of a binary classifier system[115–118]. The ROC curve has been widely used to evaluate the specificity and sensitivity of results. The Area Under the ROC Curve (AUC) provides an intuitive numerical value to measure the prediction results of the model, ranging from 0 to 1, with 1 indicating perfect prediction and 0 indicating completely wrong prediction.

## 3. Conclusion and Perspective

The integration of machine learning with natural product chemistry is not merely an enhancement of analytical capabilities but a reinvention of discovery and analysis paradigms. The potential of machine learning extends beyond incremental improvements, hinting at a revolution in our approach to understanding the molecular intricacies of natural compounds. Future developments could usher in a new era where machine learning not only assists chemists but also leads them to insights previously obscured by the sheer complexity of natural product matrices.

The advent of more sophisticated algorithms promises to refine spectral analysis to such an extent that it becomes possible to discern molecular signatures that were once indistinguishable. This precision opens the door to a deeper comprehension of biological synergies and the potential to unveil novel compounds with therapeutic benefits yet to be imagined.

Furthermore, as machine learning advances, it will become more accessible. The democratization of these tools could lead to a widespread proliferation of knowledge and an acceleration in the pace of discovery, as researchers from diverse backgrounds contribute to and benefit from advanced analytical techniques. However, this future is contingent upon responsible stewardship of machine learning technologies. Ethical application, transparency in methods, and protection of data privacy must be integral to this evolution. By adhering to these principles, the scientific community can ensure that the integration of machine learning into natural product analysis remains a beacon of progress and a testament to the collaborative spirit of scientific inquiry.

In conclusion, the intersection of machine learning and natural product analysis is poised to redefine the boundaries of what is possible in chemistry. The challenges ahead are matched only by the potential rewards: a future where natural products are understood with a clarity that accelerates innovation and enriches our collective knowledge.

## Acknowledgments
The schematic figures: "component determination", "concentration prediction" and "components classification" in Figure 1 are generated automatically from Chatgpt4.

## Abbreviations
HPLC: High-Performance Liquid Chromatography
UV-vis: Ultraviolet–visible
NMR: nuclear magnetic resonance spectroscopy
MS: mass spectrometry
IR: Infrared Radiation(IR)
ATR-FTIR: attenuated total reflection Fourier transform infrared
FTIR: Fourier transform infrared
LC-MS: Liquid chromatography-mass spectrometry
GC-MS: Gas Chromatography Mass Spectrometry
TLC:  Thin-layer chromography
MEKC: Micellar electrokinetic capillary chromatography
CF: caffeine

PCT: Paracetamol
ACT: acetosal
ML: Machine learning
EDA: Exploratory data analysis
SG-smoothing: Savitzky-Golay smoothing
PLS: Partial least squares
PCA: Principal component analysis
SVM: support vector machine
PLS-DA: Partial Lease Square-discriminant analysis
qHNMR: Quantitative 1D Proton NMR
HSQC: heteronuclear single quantum correlation
UV-Vis-NIR: Ultraviolet Visible Near Infrared Spectroscopy
ALS: Asymmetric Least Squares
SNV: Standard normal Variate
MSC: Multiplicative Scatter Correction
COW: Correlation optimized warping
PAFFT: Alignment by Fast Fourier Transform
DnCNNs: Denoising convolutional neural network
XGboost: eXtreme Gradient Boosting
MAE: Mean Absolute Error
MSE & RMSE: Mean Squared Error & Root Mean Squared Error
TP: True positives
FP: False positives
TN: True negatives
FN: False negatives
ROC curve: Receiver Operating Characteristic Curve
FPR: False positive rate
AUC curve: Area Under the ROC Curve